\documentstyle[12pt,aaspp4,epsfig]{article}

\newcommand{\fbfive}{F(162){}}
\newcommand{\mbfive}{m(162){}}

\newcommand{\ha}{H$\alpha${}}
\newcommand{\dexp}[1]{$\times 10^{#1}$}

\newcommand{\flx}{erg (cm$^{2}$\AA~s)$^{-1}$ {}}
\newcommand{\flxl}{erg (cm$^{2}$\ s)$^{-1}$ {}}

\newcommand{\surfbr}{erg (cm$^{2}$\AA~s arcsec $^{2}$)$^{-1}$ {}}

\newcommand{\hb}{H$\beta${}}
\newcommand{\ea}{{\em et~al.\ }{}}
\newcommand{\msol}{M$_\odot$}

\newcommand{\hi}{H{\small I}}
\newcommand{\hii}{H{\small II}}

\newcommand{\kms}{km~s$^{-1}$ }

\newcommand{\pasa}{{\em Pub.~Astr.~Soc.~Austr.}}

\begin{document}

\title{UIT: Ultraviolet Observations of the Small Magellanic Cloud}

\author{  Robert H. Cornett\altaffilmark{1}
           Michael R. Greason\altaffilmark{1}  
           Jesse K. Hill\altaffilmark{1}  \\
           Joel Wm. Parker\altaffilmark{2} 
           William H. Waller\altaffilmark{1} \\
          Ralph C. Bohlin\altaffilmark{3}  
          Kwang-Peng Cheng\altaffilmark{4}  
          Susan G. Neff\altaffilmark{5}
          Robert W. O'Connell\altaffilmark{6} \\ 
          Morton S. Roberts\altaffilmark{7}
         Andrew M. Smith\altaffilmark{5}  
      and Theodore P. Stecher\altaffilmark{5}  }        
          
\altaffiltext{1}{Hughes STX Corporation, Code 681, Goddard Space Flight Center,
  Greenbelt MD 20771} 
\altaffiltext{2}{Southwest Research Institute, Boulder CO} 
\altaffiltext{3}{Space Telescope Science Institute, Homewood Campus, 
 Baltimore, MD 21218}
\altaffiltext{4}{Department of Physics, California State University, 
  Fullerton, CA 92634}
\altaffiltext{5}{Laboratory for Astronomy and Solar Physics,
  Code 680, Goddard Space Flight Center,  Greenbelt MD 20771}
\altaffiltext{6}{University of Virginia, Astronomy Department,
  P.O. Box 3818, Charlottesville, VA 22903}
\altaffiltext{7}{National Radio Astronomy Observatory, Edgemont Rd.,
   Charlottesville, VA 22903}


\begin{abstract}

A mosaic of four UIT far-UV (FUV) ($\lambda_{eff}$~=~1620\AA) images, with 
derived stellar and \hii\ region photometry, is presented for most of the 
Bar of the SMC.  The UV morphology of the SMC's Bar shows that recent star 
formation there has left striking features including: a) four 
concentrations of UV-bright stars spread from northeast to southwest 
at nearly equal ($\sim$30 arcmin=0.5 kpc) spacings; b) one of the 
concentrations, near DEM 55, comprises a well-defined 8-arcmin diameter 
ring surrounded by a larger \ha\ ring, suggestive of sequential star 
formation.  

FUV PSF photometry is obtained for 11,306 stars in the FUV images, resulting
in magnitudes \mbfive.  We present a FUV luminosity function for the SMC bar, 
complete to \mbfive$\sim$14.5.  Detected objects are well correlated with 
other SMC Population I material; of 711 \ha\ emission-line stars and small
nebulae within the UIT fields of view, 520 are identified with FUV sources. 
The FUV photometry is compared with 
available ground-based catalogs of supergiants, yielding 191 detections of 
195 supergiants with spectral type earlier than F0 in the UIT fields.  The 
(\mbfive$-$V) color for supergiants is a sensitive measure of spectral type.  
The bluest observed colors for each type agree well with colors computed from 
unreddened Galactic spectral atlas stars for types earlier than about A0; for 
later spectral types the 
observed SMC stars range significantly bluer, as predicted by comparison of 
low-metallicity and Galactic-composition models.  Redder colors for some 
stars of all spectral types are attributed to the strong FUV extinction 
arising from even small amounts of SMC dust.  Internal SMC reddenings are 
determined for all catalog stars.  All stars with E(B$-$V)$>$0.15 are within 
regions of visible \ha\ emission.  
  
FUV photometry for 42 \ha-selected \hii\ regions in the SMC Bar is 
obtained for stars and for total emission (as measured in \hii-region-sized 
apertures).   The  flux-weighted average ratio of total to stellar FUV flux 
is 2.15; consideration of the stellar FUV luminosity function indicates
that most of the excess total flux is due to scattered FUV radiation, rather 
than stars fainter than \mbfive=14.5. Both stellar and total emission are well correlated with \ha\ fluxes measured 
by Kennicutt and Hodge (1986; hereafter KH), yielding FUV/\ha\ flux ratios
that are consistent with models of 
SMC metallicity, ages from 1-5 Myr, and moderate (E(B$-$V)=0.0--0.1 mag) 
internal SMC extinction.  

\end{abstract}

\section{Introduction}
UV observations from above the earth's atmosphere
are vital for understanding the Population I properties of metal-poor
galaxies such as the Small Magellanic Cloud (SMC).  Optical-band studies of
the SMC have shown that this nearby dwarf irregular galaxy provides a 
unique laboratory for investigating stellar and interstellar evolution at
``primitive'' compositions, compared to the Galaxy.  Many
effects of composition differences appear best, or only, in the UV.
Line blanketing strongly affects UV colors; similarly, the steep
SMC extinction curve (A$_{162}/E(B-V)\sim16$), widely thought to be due 
to abundances in SMC dust, is ``extreme'' only in the UV.   Moreover, because 
the energy distributions of hot stars peak at short wavelengths, FUV 
photometry is more effective than optical-band photometry in determining the 
temperatures of such stars.  Here, we present initial results from a UV 
imaging study of the the SMC, based on a mosaic of four far-UV images that
were obtained by the Ultraviolet Imaging Telescope (UIT) during the 
Astro-1 and Astro-2 missions in late 1990 and early 1995.
                              
\section{Observations and Data Reduction}

UIT observed a total of four 40-arcmin diameter fields nearly completely 
covering the SMC Bar at an angular resolution of $\sim$3 arcsec-- better than 
75 times the resolution of previous UV imaging studies of the SMC 
(\cite{okumura}). 
A description of individual exposures used in this study is in 
Table~\ref{fuvobstab}, and a mosaic of the longest exposures for each 
field is shown as Figure \ref{mosaic}. Details of UIT hardware, 
calibration, operations, and data reduction are in \cite{stechera} and 
\cite{stecherb}.  Data discussed here are from images
made with the $\lambda_{eff}$=1620\AA\ filter, hereafter called 'B5'.   
Astro-1 data also include short exposures of Field 1 made in a near-UV 
($\lambda_{eff}$=2490\AA) and an additional far-UV ($\lambda_{eff}$=1520\AA) 
bandpass (cf \cite{corn}) which are not discussed here.  (A discussion and 
analysis of UIT imaging of the Large Magellanic Cloud is given in 
\cite{parker}.)   

\placetable{fuvobstab}

UIT film images are reduced to linearized arrays and absolutely calibrated
by comparison to other UV spectrophotometry as described in \cite{stechera}.
The calibration for Astro-2 data has been revised using flight observations
of standard fields (\cite{stecherb}); estimated absolute uncertainty is 
15\% for well-exposed pixels.

An IDL/UIT implementation of DAOPHOT (\cite{stetson}; \cite{jkha})
is used to locate stars and perform aperture and PSF-fit photometry.  
Both aperture and PSF-fit photometry use a 2.9-arcsec (2.5-pixel) 
radius, which sets the limit for resolution of blended stars 
and the minimum size for non-stellar sources.  PSF-fit photometry, using 
parameters chosen for images individually, is used for subsequent analysis 
of stellar objects.  Astrometric solutions are derived for UIT images 
using HST guide stars as standards (\cite{lasker}).  11306 stars are 
measured, and positions, magnitudes, and errors for these objects are 
provided in the AAS CD-ROM series.  The file includes UIT star number, 
field number, x and y positions in UIT pixels, right ascension and 
declination, \mbfive\ derived from PSF photometry, and estimated error 
in \mbfive. 

For the SMC data, IUE observations are used to measure combined aperture 
corrections and improved single-field absolute calibrations.  The corrections 
are determined by integrating well-exposed and calibrated (Bohlin \ea 1990) 
IUE spectra of isolated field stars in each UIT image, and comparing
the results with UIT  PSF photometry. The number of IUE spectra used for 
each field ranges from 12 to 31.  Statistical uncertainties in mean 
IUE/UIT ratios range from .07 to .02 mag, with larger 
uncertainties for longer UIT exposures because of the overexposure of the 
IUE sources on those images.  The IUE/UIT ratio thus derived
agrees with the nominal UIT absolute calibration within 10\%, after application
of a directly measured aperture correction.  UIT  magnitudes are defined by   
m$_{\lambda}$~=~$-$2.5 log(F$_{\lambda}$)$-$21.1, where 
F$_{\lambda}$ is in \flx.  

Data from multiple exposures of each field are combined by selecting data
from individual exposures which maximize the signal-to-noise ratio.  
For stellar images and pixels which are common to two fields, data are
used from the field with the longer maximum exposure time, after signal-to-noise
selection. 
In practice signal-to-noise selection is achieved for stellar photometry by 
restricting each exposure to a selected stellar magnitude range; this avoids 
systematic errors at the extremes of the UIT dynamic range on single 
exposures.  For photometry of  non-stellar objects, an extended-dynamic-range, 
calibrated composite image is produced for each field by 1) boxaveraging all 
exposures to a $\sim$4.5 arcsec pixel size (to minimize the effects of possible 
densitometry misalignments of up to $\sim$1.12 arcsec), and 2) selecting and 
using pixels from the individual boxaveraged frames with exposure values 
which maximize signal-to-noise ratios.  This composite 
image is subsequently used for all extended source photometry.

\section{Far-UV Morphology}

Figure \ref{mosaic} shows that FUV emission from the SMC originates mostly in 
hot stellar populations which, while not restricted to clusters,
are significantly clumped.  No diffuse FUV emission is readily apparent (but 
see Section 5).  
The brightest features in the UIT images are the clusters NGC 346 and NGC 330, 
with additional FUV concentrations approximately centered in fields 1 and 2.  
The bright FUV concentrations, spaced along the Bar centered at 
$\sim$0.5kpc intervals, have similar clustering and distribution properties to
those evident in wide-field FUV images of the LMC (\cite{asmith}). 

\placefigure{hapic}

Comparison of large-scale \ha\  and FUV 
morphologies (Figure \ref{hapic} a) and b)); \cite{bothun}) can reveal 
spatial and temporal sequences of star formation over timescales of 
$\sim$20\ Myr (\cite{rsh}).  Ionized hydrogen emission
marks the locations of the earliest-type stars, while detectible FUV flux is 
emitted by stars as late as A0.  Therefore, \ha\ is strong for clusters up to
a few million years of age, while detectible FUV emission continues for about
20 Myr. Figures~\ref{hapic}a) and b) show that FUV concentrations cover
a range of ages, including cases with coincident bright young \ha\ 
sources at the NE and SW ends of the bar, as well as evolved regions without
\ha.

UIT field 2, near the Bar's center, provides a intriguing instance of what
appears to be sequential star formation.  A ring of FUV-bright stars with 
a radius of about 4--5 arcmin ($\sim$70--90 pc at a distance of 60 kpc; 
\cite{hutch}) dominates the field.  Figures~\ref{hapic} a) and b)
show that the stellar ring is bounded by, and is apparently interacting with, 
ionized hydrogen in a concentric, larger ring (radius  7--8 arcmin or 
$\sim$120--140pc).  Supernova remnant 0050-728 (\cite{math}), outlined by 
the circle in Figure~\ref{hapic}b, lies at
the northern edge of the shell.   \hi\ observations (\cite{staveley}, 
\cite{staveleyb}) 
show loops comparable in size to the \ha\ ring near these locations at a range 
of velocities between about 90 and 130 \kms; the ellipse in 
Figure~\ref{hapic}b shows
the location of the ring seen at $\sim$110 \kms.  Fabry-Perot \ha\ measurements 
 (\cite{okumura}) show linewidths of up to several tens of \kms, reinforcing 
evidence for ties between the distribution of hot stars and the gas dynamics.  
In a simple model, the hollow-shell morphology of the \hii\ has been driven 
by winds from now-defunct hot stars and supernova remnants, which have also 
propagated star formation from the center outward, as evidenced by relatively 
gas-free stars in the center of the \hii\ ring.  UIT-based detailed analysis of 
sequential star formation in this region will be the subject of future work 
(\cite{cornb}).  For example, comparing the distribution of UV-bright stars
with that of \ha\ emission-line stars and small nebulae as catalogued by 
(\cite{meyazz}) will reveal any systematic stellar age gradients in the 
0-30 Myr range. 

\section{Stellar Photometry}

\begin{figure}[b!] 
\setcounter{figure}{2}
\centerline{\epsfig{file=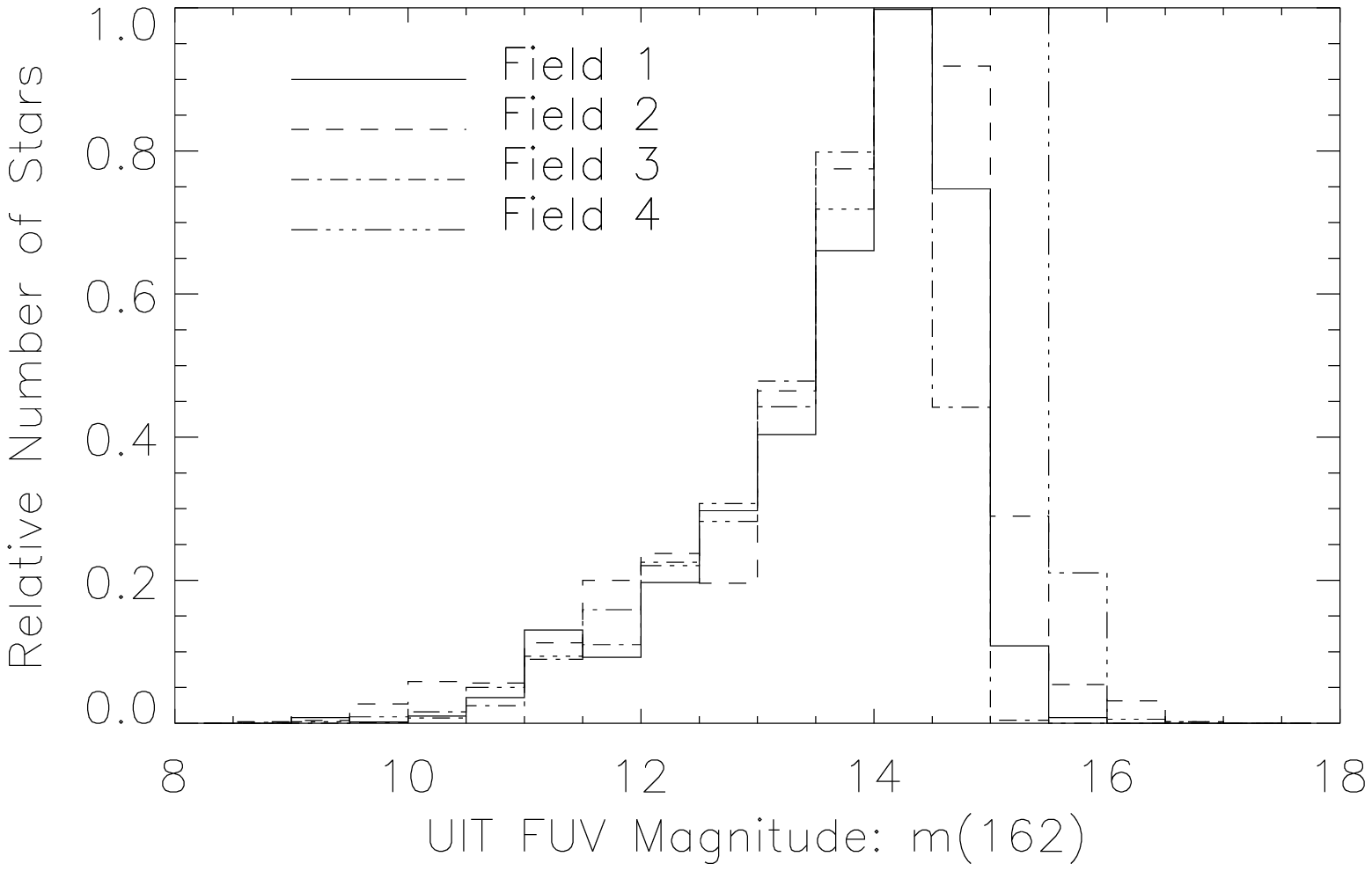,height=4.5in,width=7.in}}
\caption{\mbfive\ luminosity functions for the four fields, normalized 
to the number of stars in each for 14.0 $<$ \mbfive $<$ 14.5.
The luminosity functions for fields 1 and 2 show features that are 
confirmed by other data.  The peak for 11.0 $<$ \mbfive $<$ 11.5 in field 1's
luminosity function (solid line) is clearly displayed in the field's 
color-magnitude diagram as a supergiant ``plume'' of stars leaving the main 
sequence (\protect\cite{corn}).  In field 2 (dashed line), the overabundance 
of stars with 10.0 $<$ \mbfive $<$ 11.0 may be attributable to the effects of 
sequential star formation in the field.}

\end{figure}

Luminosity functions for the four fields, normalized to the number of stars
in each for 14.0 $<$ \mbfive $<$ 14.5, are shown in Figure 3.  From 
turnovers of the luminosity functions we estimate the limiting FUV magnitudes
of the fields to be 14.5, 14.5, 14.5, and 15.0 ($\pm\sim$0.5) for fields 1-4
respectively, consistent with relative exposure times.

Luminosity functions for fields 1 and 2 show features that may be confirmed 
from other data.  The peak for 11.0 $<$ \mbfive $<$ 11.5 in field 1's 
luminosity function (solid line) is clearly displayed in the field's 
color-magnitude diagram as a supergiant ``plume'' of stars leaving the main 
sequence (\cite{corn}).  Field 2 (dashed line) shows a significant 
overabundance of stars with 10.0 $<$\mbfive$<$ 11.0 (more than 3$\sigma$ 
overabundant compared to each of the other three fields individually) which 
may be related to sequential star formation in that field.   
The high-luminosity ends of these \mbfive\ distributions evidently 
show structure which reflects recent star formation. 

As shown in previous UIT results ({\em e.g.} \cite{jkhb}) FUV sources 
are generally well-correlated with \ha\ emission.   We have compared FUV
point-source locations with those of the 711 \ha\ emission-line stars and small
nebulae within the UIT fields of view (\cite{meyazz}), finding 520 matches 
at a matching radius of 7.2 arcsec (approximately twice the 
estimated 1-$\sigma$ the UIT astrometric uncertainty for these images).  As
described above, a detailed comparison of the spatial distribution of 
FUV sources with and without associated \ha\ emission may provide insights 
into possible sequential star formation in the SMC. 

We have also correlated our stellar photometry tables with the ground-based
stellar photometry compiled by  Azzopardi \& Vigneau (1982; hereafter AV)
by positionally matching sources.   The precision to which stellar positions 
are quoted in AV ($\sim$0.1 arcmin) does not permit unambiguous identification
of stars fainter than, typically, \mbfive$\sim$15 because several such stars
typically occur in a part of a UIT image of that size.  For AV supergiants, 
this \mbfive\ limit corresponds to a 
spectral type of $\sim$F0; therefore, we do not identify or list \mbfive\ 
values for AV stars of spectral type F0 or later.  (Where available we have 
used the spectral types listed in AV  from \cite{azo} and other sources which 
employ medium resolution slit spectra as the basis for spectral classification.)
Under this spectral type restriction we 
find 191 sources matched of the 195 within the UIT fields of view.  Results 
of the photometry of these objects are presented in Table~\ref{avtab}.

\placetable{avtab}

\begin{figure}[b!] 
\setcounter{figure}{3}
\centerline{\epsfig{file=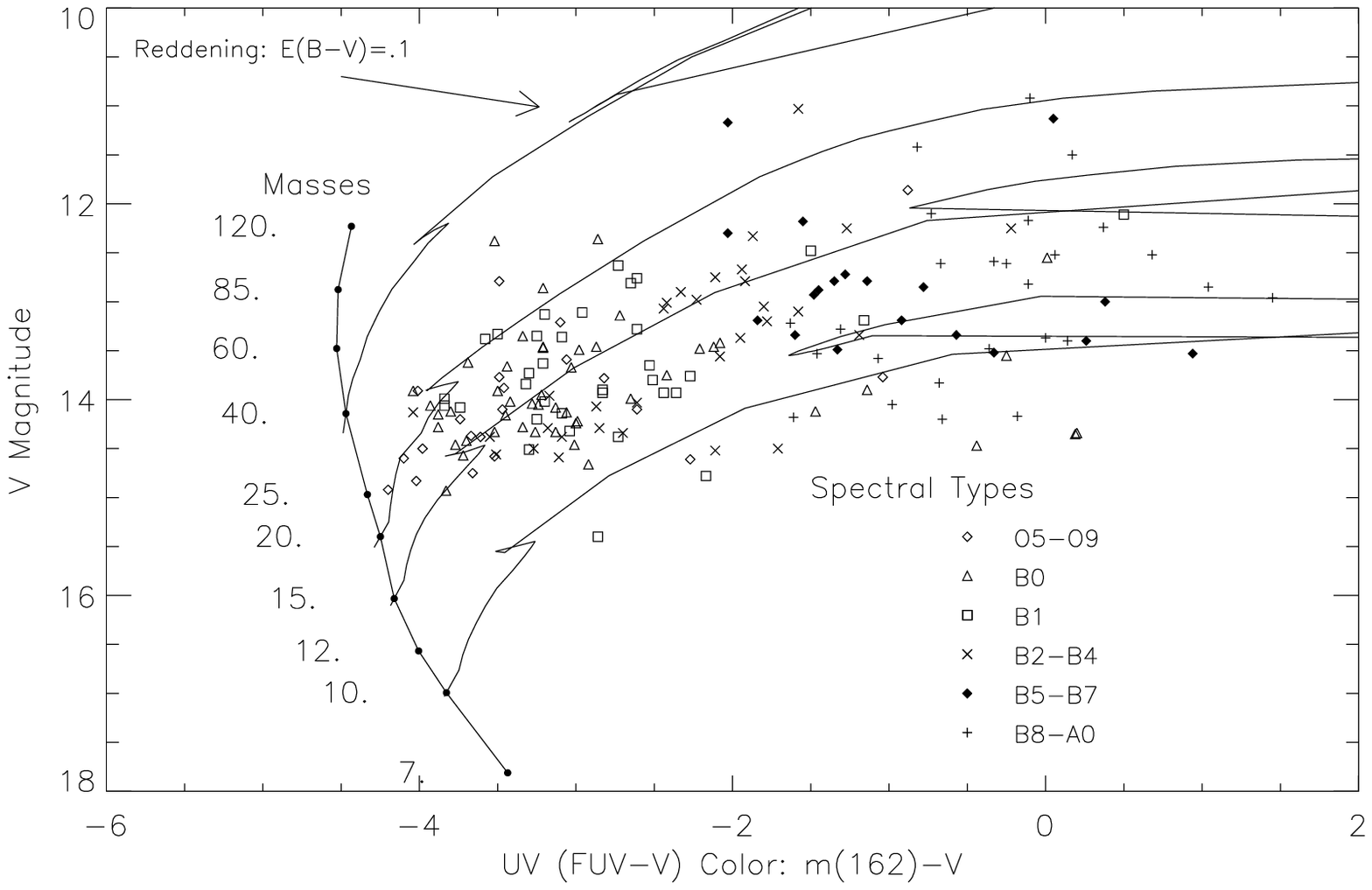,height=4.5in,width=7.in}}
\caption{The (\mbfive$-$V),V color-magnitude diagram for
191 stars in common with the catalog of Azzopardi \& Vigneau (1982).  Discrete
symbols are observed stars, with spectral types as noted in the figure.
No reddening corrections have been applied to the observed data.  The solid
lines show SMC-composition stellar models produced by using model
atmospheres of \protect\cite{kurucz} on stellar evolution models of
\protect\cite{char}. Filled circles outline a 1 Myr isochrone, with masses 
marked; the tracks to the upper right show the subsequent evolutionary paths of
10, 15, 20, and 40 \msol\  stars.  Model stars are adjusted to distance modulus
18.9 and foreground Galactic extinction E(B$-$V)=0.02 magnitudes
(\protect\cite{hutch}).  The reddening vector shown, using the SMC extinction
law of \protect\cite{hutch} and E(B$-$V)=0.1, is a typical large value
for the SMC (\protect\cite{westerlund}).   The evolutionary tracks show that
these stars
predominantly have masses between 10 and 20 \msol\  and imply ages
between 10 and 30 Myr. The lower limit corresponds to a real absence of
supergiant stars brighter than about V=11.  The general segregation of
spectral types by color implies that much of the (\mbfive$-$V) color
variation seen in this figure is due to the intrinsic colors
of the stars themselves.}

\end{figure}

Figure 4 is a color-magnitude diagram of the 191 identified 
stars common to our data and AV, with spectral classes differentiated by 
different symbols.  No reddening corrections have been applied to observed 
data.  SMC-composition stellar models are produced by using model 
atmospheres of \cite{kurucz} 
 with log(Z/Z$_\odot$)~=~$-$0.5 ($\sim$solar/3.2), on stellar 
evolution models of \cite{char} with z~=~0.004 ($\sim$solar/5.).
These grid points are selected as models with composition parameters 
nearest to the SMC's ($\sim$solar/4; \cite{westerlund}).   Filled circles 
outline a 1 Myr isochrone, with masses marked; the tracks to the upper 
right show the subsequent evolutionary paths of 10, 15, 20, and 40 \msol 
stars.  
Model stars are adjusted to distance modulus 18.9 and foreground Galactic 
extinction E(B$-$V)=0.02 magnitudes (\cite{hutch}).  The reddening vector 
shown, using the SMC extinction curve of \cite{hutch} and E(B$-$V)=0.1, is 
a typical large value for the SMC (\cite{westerlund}).  The large FUV/V 
leverage of extinction in the SMC (A$_{162}$/E(B-V)=15.75) is reflected 
in the nearly-horizontal orientation of the reddening vector. 
 
The stars catalogued by AV, the visually brightest non-cluster stars in 
the SMC, are known to be supergiants of a range of spectral types.  
The evolutionary tracks in Figure 4 clearly demonstrate that 
these stars 
predominantly have masses between 10 and 20 \msol, and therefore, ages 
between 10 and 30 Myr.  The upper age limit is set by the faintness
limit of the AV data; however, the lower limit corresponds
to a real absence of supergiant stars brighter than about V=11 
(M$\sim$30\msol.)  While 
the effects of some reddening (seen as excesses of (\mbfive$-$V) color of up 
to 1.5) are apparent, most color variation seen in this figure is due to the 
intrinsic color variations of the stars themselves, as seen from the 
general segregation of spectral types by color. 

\begin{figure}[b!] 
\setcounter{figure}{4}
\centerline{\epsfig{file=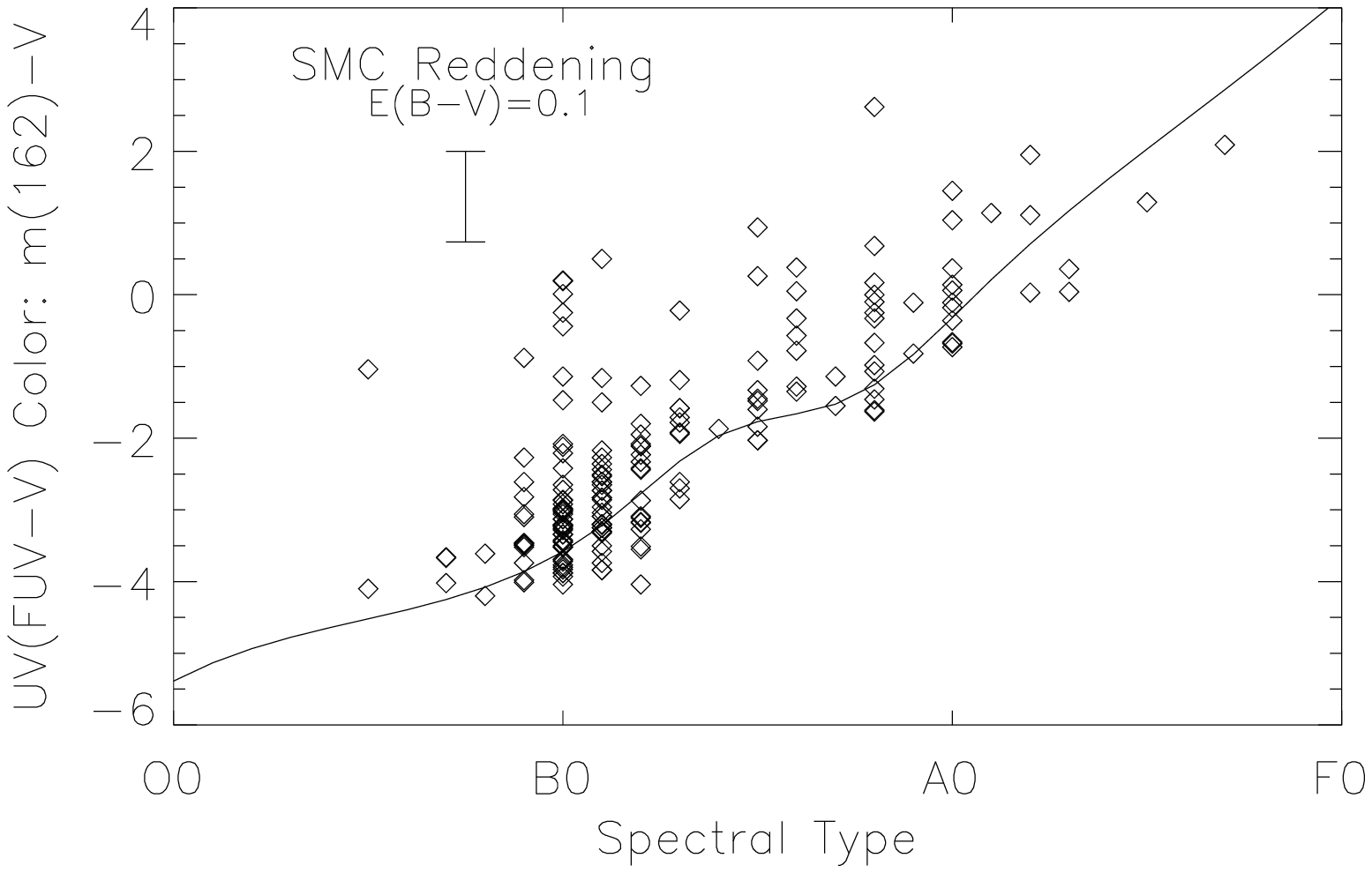,height=4.5in,width=7.in}}
\caption{{\em Diamonds:} Stellar spectral type vs (\mbfive$-$V) color for 
stars from the catalog of Azzopardi \& Vigneau (1982).  Where available, 
spectral types listed in Azzopardi \& Vigneau  from \protect\cite{azo} and 
other sources are used, which employ medium resolution slit spectra
 as the basis for spectral classification.  No reddening 
corrections have been applied to the observed color data.
{\em Solid line:} Spline fit to the (\mbfive$-$V) colors derived from the 
spectral atlas of unreddened Galactic stars of \protect\cite{fan}, reddened
by E(B$-$V)=0.02.  Galactic and SMC stellar (\mbfive$-$V) colors differ by
less than 0.10 for stars bluer than (\mbfive$-$V)=$-$1.0; SMC stars are 
significantly bluer than Galactic stars beyond that limit. } 

\end{figure}

Figure 5 illustrates the strong variation in (FUV-V) color with 
spectral type.  The data points are 
the observed colors, as in Figure 4, plotted against 
spectral type from AV.   The solid 
line is a spline fit to the observed (\mbfive$-$V) colors of Galactic 
supergiants from the atlas of \cite{fan}. The atlas colors have been reddened 
by the amount of the SMC's Galactic foreground reddening.  The sensitivity of 
the (\mbfive$-$V) color to spectral type is evident, as are the significant 
reddenings for some stars, caused by the large FUV leverage of the SMC 
extinction law.  The atlas colors 
are seen to form a blue limit for the observed colors, as expected for 
unreddened stars. 

While a true calibration of color {\em versus} spectral type for SMC stars 
is not available, Galactic values are probably appropriate for blue stars: 
(\mbfive$-$V) colors of Galactic- and SMC-composition model atmospheres 
differ by less than 0.10 for models that have 
(\mbfive$-$V)$<-$1.0.  However, at Galactic (\mbfive$-$V)=1.5, SMC models are 
approximately 0.5 mag bluer, with larger differences for redder stars.  
Therefore, much of the apparent discrepancy between the blue limit of the 
observed colors and the atlas colors is probably due to the fact that the 
locus of colors from an ``SMC atlas'' would be significantly 
bluer for stars later than about A0, than the Galactic atlas colors shown.  

Figure 5 provides a straightforward method of computing the 
reddenings along the lines of sight to the stars in AV. We do so by 
comparing the observed colors to atlas colors for the spectral type, 
corrected for Galactic foreground reddening, and converting the difference 
to E(B$-$V) using the extinction law of \cite{hutch}.  Extinctions computed in 
this way are presented in Table~\ref{avtab} for all identified AV stars.  Stars 
later in spectral type than B8, which have uncertain intrinsic colors, 
are noted.   

\placefigure{redmap}

Figure \ref{redmap} shows the resulting E(B$-$V) values plotted at the 
locations of the stars on the SMC, against the \ha\ image shown in 
Figure~\ref{hapic}a.
All values of E(B$-$V) $>\sim$0.15 may be seen
from Figure~\ref{hapic}b to be at least along the line of sight to \ha\ emission.  
The correlation of large reddening values with location implies that these 
stars are probably associated with the nebulosity. 

In the future UIT photometry of the SMC will be supplemented by 
additional ground-based observations currently underway by the UIT team, as well
as by the comprehensive photometric surveys of the Magellanic Clouds
(\cite{zar}).

\section{\hii~ Region Photometry} 

\ha\ emission from \hii\ regions indirectly traces the OB star content of the 
regions via the process of nebular ionization by the hot stars and subsequent 
recombination of the hydrogen gas.  The FUV continuum emission ({\em e.g.} 
at 1620\AA) from \hii\ regions arises from OB stars as well as from the 
more numerous non-ionizing stars (mid-B through early A types.) 
Therefore, the comparison of \ha\ and FUV fluxes from \hii\ regions provides
a sensitive measure of each \hii\ region's evolutionary state. Evolutionary 
models of single-burst ionizing clusters confirm that the 
ratio of Lyman continuum to FUV flux rapidly decreases with age, with the 
FUV emission remaining significant for $\sim$20 Myr (\cite{rsh}). 

We have measured FUV fluxes for 42 \ha-bright \hii\ regions from the catalog of
Davies, Elliott, and Meaburn 1976 (hereafter \cite{dem}) that were measured
at \ha\ by Kennicutt \& Hodge 1986 (hereafter \cite{kh}), and have compared
the observed flux ratios with predictions from cluster models.  We compute 
FUV fluxes in two ways: by summing the flux from stars in the apertures 
defined by KH; and by summing all flux in pixels 
in the apertures-- the latter method including diffuse FUV emission.  
(Our use of the KH aperture definitions causes significant 
aperture overlaps in a few cases.)  Table~\ref{hiitab} presents fluxes and errors: 
\ha\ from
KH; FUV from stars (from PSF fits to stars in the aperture);
and FUV total fluxes from the apertures themselves.  

\placetable{hiitab} 

For the FUV aperture measurements the ``sky'' flux has been set to zero, 
and the error term stated does not include a sky contribution.  We choose 
this approximation because the actual sky contribution to any FUV aperture 
measurement or its uncertainty is very small, and any aperture 
inside the UIT image area will measure a high background due to SMC material.
We set limits on true sky flux by measuring the average background in Astro-2 
images made in the UIT B5 filter of a nearby field containing part of the 
globular cluster 47 Tuc.  These images, long exposures (1030 and 1800 sec) made 
during orbital day, have average earth limb elevation angles and other 
observational conditions which are typical of those during UIT SMC 
observations, and therefore give representative background measurements 
(\cite{waller}).  As measured in this way the sky contribution to the
FUV aperture fluxes is 3.3\dexp{-18}\surfbr, with an average contribution of 
0.75\% and a maximum of 1.60\% to the individual aperture fluxes.  These 
estimates of ``sky'' contributions are also listed in Table~\ref{hiitab}.

\begin{figure}[b!] 
\setcounter{figure}{6}

\centerline{\epsfig{file=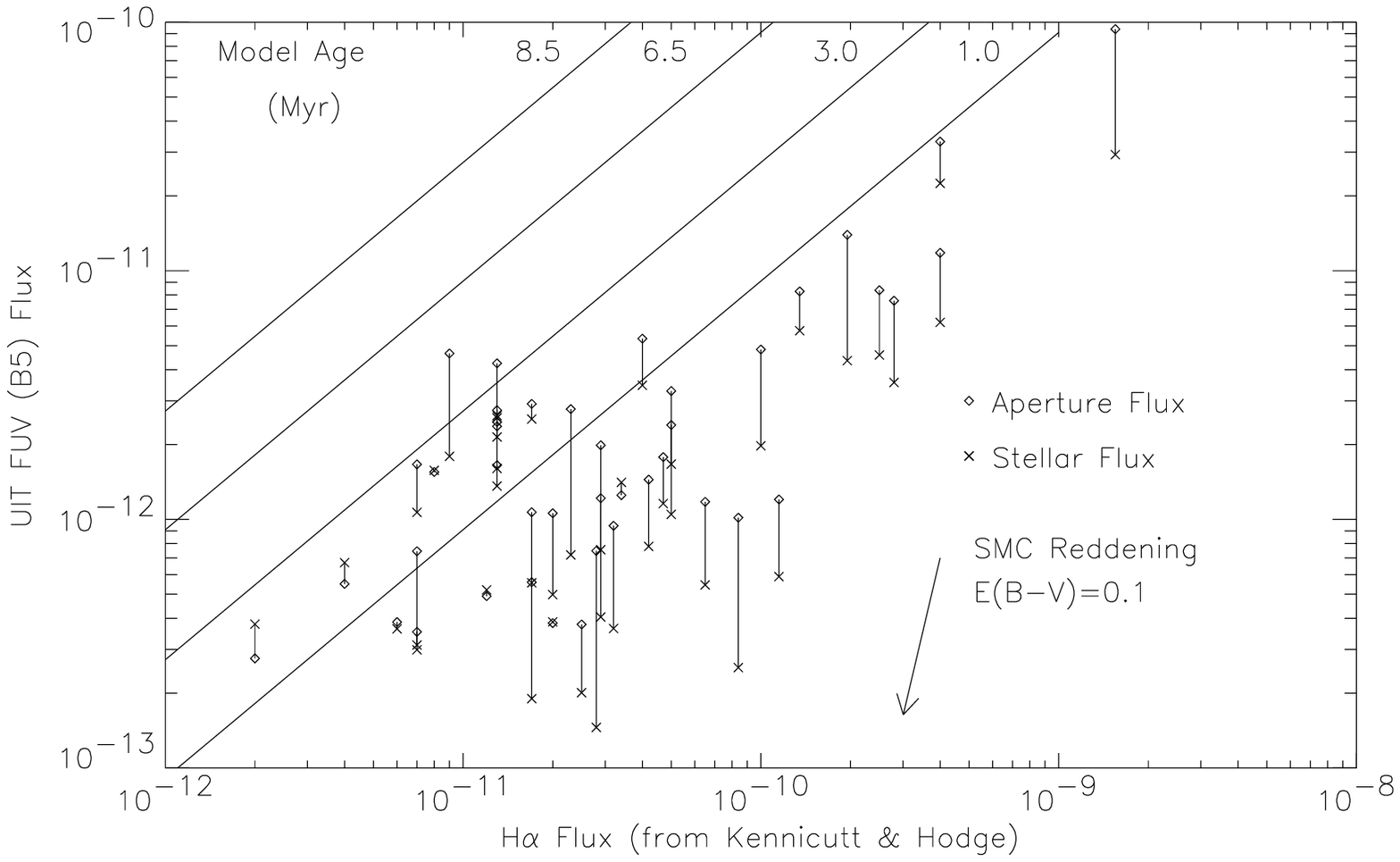,height=4.5in,width=7.in}}
\caption{FUV ($\lambda_{eff}$~=~1620\AA) vs \ha\ flux for \hii\ regions as
defined by the apertures of Kennicutt \& Hodge (1986). {\em Crosses}
mark the total FUV stellar flux, from our stellar photometry, of stars which
lie within the KH apertures.  {\em Diamonds} mark the total FUV flux contained
within the apertures.  No reddening corrections are applied to the data,
although a typical SMC internal reddening corresponding to E(B$-$V)=0.1
is shown.  {\em Solid lines} at constant FUV/\ha\ mark the expected
FUV/\ha\ ratios for clusters of single-burst formation of stars of SMC
composition, reddened to Galactic foreground values.  The ages of the clusters
are 1 Myr, 3 Myr, 6.5 Myr, and 8.5 Myr respectively, with the smallest FUV/\ha\
value corresponding to age 1 Myr.}
\end{figure}

Figure 7 shows the data from Table~\ref{hiitab}. Open diamonds are 
aperture fluxes, 
and crosses are stellar photometry.  No corrections for Galactic 
foreground reddening have been made to the data.  The FUV flux is generally 
well correlated with its \ha\ counterpart; furthermore, the ratio of aperture 
to stellar FUV flux is relatively uniform, especially for brighter 
\hii\ regions.  This ratio measures the relative amounts of bright-star and
``diffuse'' light, where ``diffuse'' light here includes contributions from
faint stars as well as from dust scattering.  The mean value of the ratio of 
aperture flux to stellar flux, averaged in the log ratio for all \hii\ 
regions, is 1.79; weighted by flux (the total aperture flux for all 
\hii\ regions divided by the total stellar flux for all \hii\
regions) is 2.15.  Two lines of argument point to a dust-scattering origin
for most of the additional FUV flux, however.  First, \cite{bohlin} 
have determined the total-to-stellar flux ratio for the Orion nebula, for which 
the number of undetected stars is small, to be 2.5 for 200-300\AA\ 
bandpasses near 1400, 1823, 2242, and 2622\AA.  Second, extrapolating our
measured luminosity function for the Field 4, which has the deepest exposure,
to \mbfive=18.0 (corresponding to an A0 main sequence star, at which 
spectral type the FUV flux drops rapidly) predicts an additional stellar 
flux contribution of 22\% and a resulting flux-weighted total-to-stellar
ratio of 1.76-- significantly smaller than the observed value. 
Scattered-light fractions near the observed SMC value are found in the 
giant \hii\ regions of M33 (\cite{malumuth}).  Although the small SMC dust 
abundance (\cite{westerlund}) undoubtedly causes the scattering to be 
significantly less than for Orion, it is nonetheless clear that 
dust-scattered radiation is a major contributor to the FUV aperture 
emission from \hii\ regions. 

Figure 7 also displays the FUV/\ha\ ratio,
known to be a good diagnostic of \hii\ region evolution (\cite{rsh}).  
We have computed \fbfive/\ha\ ratio values for clusters with a single 
burst of star formation of SMC composition using the IDL procedure CLUSTFLUX 
(\cite{landsman}).  The ratio rises essentially 
monotonically from 0.1 at 1 Myr to 4.3 at 10 Myr; ratios for cluster 
ages of 1Myr, 3Myr, 6.5 Myr, and 8.5 Myr, corrected for foreground 
Galactic reddening (which increases all \fbfive/\ha\ values by 10\%) are 
plotted as solid lines in Figure 7.  For reference, the arrow in 
Figure 7 shows a typical internal SMC extinction of E(B$-$V)=0.1;
\cite{capde} find reddenings local to SMC bar \hii\ regions in the range
0.08$<$E(B$-$V)$<$0.27 by comparing 6-cm radio, \ha, \hb, and
nearby stellar measurements.
Therefore, model and observed ratios agree well for \hii-region 
ages of a few Myr and reasonable internal SMC extinction; the observed
\hii\ regions are evidently no older than $\sim$5 Myr for any internal 
SMC reddening, and, for typical assumed SMC reddening, are between 1 Myr 
and 3 Myr in age.  This result is consistent with observations of \hii\ 
regions in galaxies as disparate as NGC 4449 (\cite{rsh}) and 
M81 (\cite{jkhb}).  There is also a clear tendency for \ha-bright 
\hii\ regions to be either relatively young-- which is not surprising-- 
or unreddened.  The absence of significantly older \hii\ regions is most 
likely a selection effect originating in KH's bias for \ha-bright \hii\
regions.  
 
\section{Summary}

UIT FUV images and
derived stellar photometry for most of the Bar of the SMC are presented.  The 
UV morphology of the SMC's Bar shows four concentrations of UV-bright stars 
spread from northeast to southwest at nearly equal ($\sim$30 arcmin) 
spacings.  One of the concentrations, near DEM 55, appears as a 
well-defined 8-arcmin diameter ring surrounded by a larger \ha\ ring
and strongly suggests sequential star formation.  

FUV PSF photometry, resulting in \mbfive\ magnitudes, is obtained for 
11,306 stars. We 
present a FUV luminosity function for the SMC bar, complete to \mbfive\  
$\sim$14.5, and compare our photometry with the compiled ground-based
data of AV. Detected objects are well correlated with
other SMC Population I material; 520 of 711 \ha\ emission-line stars and small
nebulae within the UIT fields of view are identified with FUV sources.
For early type stars, the bluest observed (\mbfive$-$V) 
colors for each spectral type agree well with values computed 
from unreddened 
Galactic spectral atlas stars for types earlier than about A0; for later 
types, observed SMC stars range significantly bluer, as predicted by 
low-metallicity models.  We attribute redder colors for some stars of 
all spectral types to strong FUV extinction due to even small amounts 
of SMC dust.  Internal SMC reddenings are determined for all catalog stars.
All stars with E(B$-$V)$>$0.15 are within regions of visible \ha\ emission.
  
FUV photometry is obtained for the resolved stars and for the total emission 
from DEM \hii\ regions in the SMC Bar.  The flux-weighted ratio of total 
to stellar flux for DEM \hii\
regions using the apertures of KH is 2.15; since only 22\% more flux is 
contributed by stars fainter than \mbfive=14.5, most of the excess 
total flux is due to scattered FUV from dust.  Stellar and total emission 
from DEM \hii\ regions are 
well correlated with \ha\ fluxes measured by KH. 
We compute ratios of FUV to \ha\ flux for 42 SMC \hii\ regions and compare
them with model results, finding that the observed ratios for all 
\ha-selected \hii\ regions are consistent with models of SMC metallicity, 
ages from 1-5 Myr, and moderate (E(B-V)=0.0-0.1 mag) extinction.  

\acknowledgements 

We gratefully acknowledge the contributions made by the many people 
involved in the {\em Astro-2} mission.  We thank Robert S. Hill, 
Wayne Landsman, and Michael Fanelli for useful discussions, Greg Bothun
for the use of his \ha\ image of the SMC, Joel Offenberg for help in 
preparing publication-quality images, and Joan Hollis and Emily Zamkoff 
for data entry and programming assistance.  We also thank Lister Staveley-Smith
for kindly supplying a digital version of a part of his \hi~ maps of the 
complete SMC.
  
Funding for the UIT project has been through the Spacelab Office at NASA
under project number 440-51.  RWO gratefully acknowledges NASA support of
portions of this research through grants NAG5-700 and NAGW-2596 to the 
University of Virginia.

\newpage

\setcounter{figure}{0}

\figcaption[cornett.fig1.ps]{A far-UV (FUV) mosaic of the Small Magellanic 
Cloud.  It is made up of the longest FUV ($\lambda_{eff}$~=~1620\AA)
exposures of SMC fields made by UIT during the Astro-1 and Astro-2 missions.
Exposure times range from 117 to 898 seconds; see Table \ref{fuvobstab} 
for observational details.  The bright clusters centered in Fields 4 and 3 
respectively are NGC 346 and NGC 330; the field of view shown encompasses 
nearly the entire SMC bar, and the resolution is $\sim$3\arcsec. \label{mosaic}}

\figcaption[cornett.fig2a.ps] {a) The SMC \ha\ "parking lot camera" image 
obtained by Bothun (personal communication) displayed on a linear greyscale.  
---b)  Linearly spaced contours from Figure 2a) superimposed on the image of 
Figure \ref{mosaic}. A constant background has been subtracted.  Note the 
prominent ring of
\ha\ emission surrounding, and larger than, the ring of FUV-bright stars 
in Field 2. SNR 0050-728 (\protect\cite{math}) is marked in size and position
by the circle at the northern edge of the ring. A prominent ring of \hi\
emission, seen over a range of radial velocities centered at $\sim$110\kms in 
the data of \protect\cite{staveley}, is outlined by the ellipse. \label{hapic}}

\setcounter{figure}{5}

\figcaption[cornett.fig6.ps]{Stellar reddening (E(B$-$V)) map, computed 
by comparing the 
observed (\mbfive$-$V) colors for stars from the catalog of Azzopardi \& 
Vigneau (1982) with the Galactic atlas spline 
fit for that spectral type shown in Figure 5. Stellar 
symbol diameter is proportional to E(B$-$V)
as shown in the legend. Observed colors have been corrected for Galactic 
foreground reddening.  All values of E(B$-$V) $>\sim$0.15 may be seen
from Figure 2b to be at least along the line of sight to \ha\ emission.  
The correlation of large reddening values with location implies that these 
stars are probably associated with the nebulosity.\label{redmap}}


\clearpage

{\sc TABLE} 1. Observational parameters for UIT images of the Small Magellanic
Cloud.

{\sc TABLE} 2. Observational data for stars from Azzopardi \& Vigneau 
1982 (AV) which were observed by UIT.  AV stars which are within UIT fields 
but later in spectral type than A9 are omitted, since they are too faint
in the FUV for detection by UIT. \mbfive values are derived from PSF 
photometry as described in the text.  E(B$-$V) values are derived
by comparing the unreddened (\mbfive$-$V) color for the stellar spectral type
with the observed color, using the reddening law of \protect\cite{hutch}.    

{\sc TABLE} 3. Fluxes for  \hii\ regions from the list of Davies, Elliott, \&
Meaburn 1976 (DEM) as measured by Kennicutt \& Hodge 1986 (KH) which are 
completely contained within the UIT field of view.  ``FUV Stellar'' flux is
the total flux of stars within the aperture defined by KH; ``FUV Aperture'' 
flux is the total pixel flux contained within the aperture defined by KH; 
and ``Sky Flux'' is an estimate for the non-SMC sky contribution within
the aperture, as described in the text.

\begin{deluxetable}{lclllll}
\tablecolumns{7}
\tablecaption{UIT FUV Images of the SMC \label{fuvobstab}}

\tablehead{
\colhead{Field}  &  \colhead{Field Center} & \colhead{Exposure Times} &
\multicolumn{2}{c}{Obs. Epoch} &
\colhead{No.}  &  \colhead{Flight} \\
\colhead{} & \colhead{RA (2000.0) Dec} & \colhead{(sec)} &
\multicolumn{2}{c}{(GMT)} & \colhead{Stars} \\ }

\startdata

  1 & 00 47 33.7  -73 06 26.8 & 23.9,117.0 & 12/05/90 & 06:16 & 1713 & Astro-1   \nl
  2 & 00 50 58.3  -72 44 56.5 & 25.9,247.0 & 03/16/95 & 23:50 & 2110 &  Astro-2  \nl
  3 & 00 56 41.2  -72 28 28.8 & 18.7,459.0 & 03/06/95 & 06:14 & 2517 &  Astro-2  \nl
  4 & 00 59 24.2  -72 11 08.4 & 36.2,179.0,898. & 03/09/95 & 04:04 & 4966 & Astro-2 \nl

\enddata

\end{deluxetable}

\begin{deluxetable}{rrrclcrrrr}
\tablecolumns{10}
\tablecaption{UIT FUV Magnitudes of AV Stars \label{avtab}}

\tablehead{
\colhead{\#} & \colhead{AV} & \colhead{UIT} &
\multicolumn{1}{c}{R.A. (2000.0)  Dec} &
\colhead{Sp} & \colhead{V} &
\multicolumn{2}{c}{UIT} &
\colhead{E(B-V)} &
\colhead{} \\
\colhead{} & \colhead {\#} & \colhead {\#} &
\multicolumn{3}{c}{(Azzopardi and Vigneau)} &
\colhead{\mbfive} & \colhead{$\sigma$} & \\ }

\startdata
   1 &  2B & 1686 & 00 43 51.2  -73 08 54 &  B0 & 13.49 & 10.51 & 0.10 & 0.05 &   \nl
   2 &   3 &   70 & 00 44 29.6  -72 58 00 &  B0 & 14.12 & 12.65 & 0.07 & 0.17 &   \nl
   3 &   6 & 1699 & 00 45 19.8  -73 15 18 &  B0 & 13.46 & 10.25 & 0.08 & 0.03 &   \nl
   4 &   7 & 1417 & 00 45 33.7  -73 04 31 &  B0 & 14.57 & 10.85 & 0.06 &-0.01 &   \nl
   5 &   8 & 1505 & 00 45 36.2  -72 59 12 &  B8 & 13.53 & 12.07 & 0.05 &-0.02 &   \nl
   6 &   9 & 1694 & 00 45 37.1  -73 14 07 &  B2 & 13.05 & 11.25 & 0.07 & 0.08 &   \nl
   7 &  11 & 1622 & 00 45 54.3  -73 16 13 &  B2 & 13.56 & 11.48 & 0.05 & 0.05 &   \nl
   8 &  12 & 1429 & 00 46  2.8  -73 06 19 &  O9 & 13.21 & 10.11 & 0.07 & 0.06 &   \nl
   9 &  14 &  321 & 00 46 33.2  -73 05 55 &  O5 & 13.77 & 12.73 & 0.07 & 0.28 &   \nl
  10 &  16 & 1558 & 00 46 55.8  -73 08 25 &  B1 & 13.19 & 12.03 & 0.10 & 0.16 &   \nl
  11 &  17 & 1522 & 00 47  3.6  -73 05 55 &  B1 & 13.76 & 11.49 & 0.12 & 0.08 &   \nl
  12 & 17A & 1679 & 00 47  4.8  -73 06 19 &  B2 & 13.37 & 11.42 & 0.10 & 0.07 &   \nl
  13 &  18 & 1433 & 00 47 13.1  -73 06 25 &  B1 & 12.48 & 10.98 & 0.04 & 0.14 &   \nl
  14 &  19 & 2766 & 00 47 15.8  -72 49 55 &  B6 & 13.00 & 13.38 & 0.08 & 0.16 &   \nl
  15 &  20 &  158 & 00 47 29.9  -73 01 25 & B8: & 12.14 & 14.76 & 0.08 & 0.31 &   \nl
  16 &  21 & 1690 & 00 47 32.6  -73 10 56 &  B1 & 14.14 & 11.05 & 0.05 & 0.01 &   \nl
  17 &  22 & 1444 & 00 47 39.8  -73 07 38 &  B2 & 12.25 & 10.98 & 0.06 & 0.12 &   \nl
  18 &  23 & 1670 & 00 47 40.1  -73 22 43 &  B3 & 12.25 & 12.03 & 0.08 & 0.17 &   \nl
  19 &  24 & 1675 & 00 47 42.9  -73 02 25 &  O9 & 13.78 & 10.96 & 0.07 & 0.08 &   \nl
  20 &  25 & 1597 & 00 47 47.5  -73 12 19 &  B5 & 13.19 & 11.35 & 0.10 &-0.01 &   \nl
  21 &  26 &  514 & 00 47 50.5  -73 08 14 &  B0 & 12.55 & 12.56 & 0.07 & 0.28 &   \nl
  22 &  27 &  947 & 00 47 51.6  -73 14 14 &  A2 & 12.17 & 14.12 & 0.05 & 0.10\tablenotemark{c} & \nl
  23 &  28 & 1662 & 00 47 55.0  -73 21 14 &  B0 & 13.42 & 11.34 & 0.08 & 0.12 &   \nl
  24 &  31 &  361 & 00 48  7.2  -73 06 32 & B8: & 12.52 & 13.20 & 0.06 & 0.15 &   \nl
  25 &  32 & 1939 & 00 48 10.2  -72 43 44 &  B1 & 14.20 & 10.95 & 0.03 & 0.00 &   \nl
  26 &  34 &   -1 & 00 48 18.7  -73 23 38 & B2: & 13.81 &  0.00 & 0.00 &-9.99\tablenotemark{b} & \nl
  27 &  35 &   -1 & 00 48 21.3  -72 51 02 & B0: & 14.13 &  0.00 & 0.00 &-9.99\tablenotemark{a} & \nl
  28 &  36 & 1934 & 00 48 23.9  -72 43 38 &  B1 & 14.02 & 10.82 & 0.04 & 0.00 &   \nl
  29 &  37 & 1676 & 00 48 27.0  -73 03 20 &  B5 & 12.88 & 11.43 & 0.07 & 0.03 &   \nl
  30 &  38 &  142 & 00 48 28.2  -73 00 32 &  A0 & 12.85 & 13.89 & 0.05 & 0.11\tablenotemark{c} & \nl
  31 & 39A & 1476 & 00 48 31.4  -73 15 32 &  W6 & 99.99 & 10.94 & 0.06 &-9.99\tablenotemark{d} & \nl
  32 &  41 & 1767 & 00 48 36.1  -72 52 50 &  B2 & 14.56 & 11.05 & 0.06 &-0.06 &   \nl
  33 &  42 & 2265 & 00 48 40.2  -72 58 14 &  B5 & 13.49 & 12.16 & 0.09 & 0.03 &   \nl
  34 &  43 & 1875 & 00 48 48.8  -72 46 14 &  B1 & 14.08 & 10.34 & 0.04 &-0.04 &   \nl
  35 &  45 &   -1 & 00 48 50.3  -73 22 02 &  B8 & 14.15 &  0.00 & 0.00 &-9.99\tablenotemark{a} & \nl
  36 &  48 & 1709 & 00 49  3.4  -73 21 32 &  B3 & 11.03 &  9.45 & 0.12 & 0.06 &   \nl
  37 &  50 & 1765 & 00 49 16.7  -72 52 32 &  B1 & 13.11 & 10.15 & 0.05 & 0.02 &   \nl
  38 &  51 & 1777 & 00 49 32.8  -72 51 08 &  B0 & 14.05 & 10.81 & 0.03 & 0.03 &   \nl
  39 &  53 & 2495 & 00 49 45.8  -72 52 38 &  A0 & 12.96 & 14.41 & 0.05 & 0.14\tablenotemark{c} & \nl
  40 &  55 & 1197 & 00 49 48.2  -73 17 45 &  B5 & 13.40 & 13.66 & 0.06 & 0.16 &   \nl
  41 &  56 & 1751 & 00 49 51.7  -72 55 39 &  B5 & 11.17 &  9.14 & 0.04 &-0.02 &   \nl
  42 &  58 & 1769 & 00 49 58.3  -72 51 45 &  B2 & 14.38 & 10.83 & 0.04 &-0.06 &   \nl
  43 &  59 &  788 & 00 49 58.4  -73 11 33 &  A0 & 13.40 & 13.54 & 0.06 & 0.04\tablenotemark{c} & \nl
  44 &  62 & 2361 & 00 50  1.2  -72 55 03 &  B3 & 14.34 & 11.64 & 0.03 &-0.03 &   \nl
  45 &  63 &  672 & 00 50  0.9  -73 10 08 &  A0 & 13.48 & 13.12 & 0.06 & 0.00\tablenotemark{c} & \nl
  46 &  65 & 1448 & 00 50  6.9  -73 07 39 &  B6 & 11.13 & 11.18 & 0.08 & 0.14 &   \nl
  47 &  66 & 1642 & 00 50  6.7  -73 16 21 &  B0 & 13.48 & 11.27 & 0.09 & 0.11 &   \nl
  48 &  67 & 2039 & 00 50 11.9  -72 32 27 &  B0 & 13.66 & 10.22 & 0.07 & 0.01 &   \nl
  49 &  69 & 2442 & 00 50 18.0  -72 53 21 &  B0 & 13.35 & 10.01 & 0.21 & 0.02 &   \nl
  50 &  70 & 2007 & 00 50 18.9  -72 38 03 &  B0 & 12.38 &  8.86 & 0.04 & 0.00 &   \nl
  51 &  73 & 1677 & 00 50 28.4  -73 03 09 &  B0 & 14.08 & 10.95 & 0.07 & 0.04 &   \nl
  52 &  75 & 1757 & 00 50 32.9  -72 52 27 &  O9 & 12.79 &  9.30 & 0.04 & 0.03 &   \nl
  53 &  77 & 1827 & 00 50 34.2  -72 47 39 &  B0 & 13.91 & 10.41 & 0.03 & 0.01 &   \nl
  54 &  80 & 1824 & 00 50 44.3  -72 47 33 &  B1 & 13.33 &  9.83 & 0.04 &-0.02 &   \nl
  55 &  82 & 1925 & 00 50 49.8  -72 42 33 &  B2 & 14.13 & 10.09 & 0.06 &-0.10 &   \nl
  56 &  83 & 1930 & 00 50 52.8  -72 42 03 &  B1 & 13.38 &  9.80 & 0.04 &-0.03 &   \nl
  57 &  85 & 2440 & 00 51  0.8  -72 52 57 &  B0 & 13.75 & 11.33 & 0.05 & 0.09 &   \nl
  58 &  87 & 3400 & 00 51  8.3  -72 40 03 &  B0 & 13.90 & 12.76 & 0.11 & 0.19 &   \nl
  59 &  89 & 1121 & 00 51 10.4  -73 15 45 &  B0 & 14.47 & 14.03 & 0.09 & 0.25 &   \nl
  60 &  90 &   -1 & 00 51 12.1  -72 28 09 &  A5 & 12.58 &  0.00 & 0.00 &-9.99\tablenotemark{a} & \nl
  61 &  91 & 2325 & 00 51 11.5  -72 55 09 &  B8 & 12.61 & 12.36 & 0.04 & 0.08 &   \nl
  62 &  93 & 1940 & 00 51 20.8  -72 41 21 & B0: & 14.13 & 11.07 & 0.06 & 0.04 &   \nl
  63 &  94 & 1783 & 00 51 20.6  -72 49 33 &  B1 & 13.99 & 10.15 & 0.04 &-0.05 &   \nl
  64 &  95 & 1885 & 00 51 21.9  -72 44 03 &  O9 & 13.91 &  9.90 & 0.04 &-0.01 &   \nl
  65 &  99 & 1809 & 00 51 25.4  -72 47 45 &  B2 & 13.01 & 10.59 & 0.04 & 0.03 &   \nl
  66 & 100 & 1743 & 00 51 26.9  -72 57 45 &  B3 & 14.29 & 11.44 & 0.05 &-0.04 &   \nl
  67 & 102 & 2036 & 00 51 36.5  -72 32 09 &  B2 & 14.29 & 11.11 & 0.07 &-0.03 &   \nl
  68 & 103 & 1847 & 00 51 36.7  -72 45 57 &  B1 & 13.36 & 10.27 & 0.03 & 0.01 &   \nl
  69 & 104 & 1800 & 00 51 39.0  -72 47 57 &  B1 & 13.13 &  9.93 & 0.04 & 0.00 &   \nl
  70 & 105 & 3885 & 00 51 41.9  -72 27 57 &  A0 & 12.24 & 12.61 & 0.07 & 0.05\tablenotemark{c} & \nl
  71 & 106 & 2005 & 00 51 44.1  -72 37 15 &  B1 & 14.32 & 11.28 & 0.05 & 0.01 &   \nl
  72 & 109 & 1976 & 00 51 50.6  -72 39 15 &  B1 & 13.73 & 10.43 & 0.03 &-0.01 &   \nl
  73 & 110 & 1877 & 00 51 52.8  -72 44 03 &  A0 & 12.17 & 12.06 & 0.06 & 0.02\tablenotemark{c} & \nl
  74 & 111 & 2035 & 00 51 56.7  -72 31 57 &  B1 & 13.84 & 10.52 & 0.06 &-0.01 &   \nl
  75 & 112 & 2031 & 00 51 58.5  -72 33 15 &  B0 & 14.15 & 10.27 & 0.05 &-0.02 &   \nl
  76 & 114 & 1971 & 00 52  3.1  -72 39 22 &  B0 & 14.93 & 11.10 & 0.05 &-0.02 &   \nl
  77 & 122 & 1932 & 00 52 25.7  -72 40 52 &  B6 & 12.79 & 11.44 & 0.05 & 0.02 &   \nl
  78 & 123 & 1758 & 00 52 27.8  -72 50 58 &  B8 & 13.22 & 11.59 & 0.06 &-0.03 &   \nl
  79 & 126 & 1961 & 00 52 31.7  -72 39 22 &  B0 & 13.47 & 10.26 & 0.04 & 0.03 &   \nl
  80 & 131 & 1841 & 00 52 41.0  -72 45 10 &  B8 & 12.61 & 11.94 & 0.07 & 0.05 &   \nl
  81 & 133 & 2003 & 00 52 44.2  -72 36 46 &  B0 & 13.91 &  9.87 & 0.04 &-0.04 &   \nl
  82 & 137 & 1865 & 00 52 53.0  -72 44 04 &  B4 & 12.33 & 10.46 & 0.04 & 0.01 &   \nl
  83 & 138 & 1784 & 00 52 52.9  -72 48 22 &  B0 & 14.28 & 10.94 & 0.05 & 0.02 &   \nl
  84 & 141 & 3723 & 00 53 10.5  -72 33 40 & B3: & 14.50 & 12.79 & 0.04 & 0.05 &   \nl
  85 & 143 & 1970 & 00 53 27.1  -72 38 22 &  B0 & 14.12 & 10.32 & 0.04 &-0.02 &   \nl
  86 & 144 & 1742 & 00 53 33.7  -72 56 16 &  B0 & 14.06 & 10.13 & 0.08 &-0.03 &   \nl
  87 & 145 & 1983 & 00 53 36.0  -72 37 40 &  B1 & 13.35 & 10.10 & 0.05 & 0.00 &   \nl
  88 & 148 & 1882 & 00 53 42.5  -72 42 28 &  B0 & 14.28 & 10.40 & 0.04 &-0.02 &   \nl
  89 & 149 & 1774 & 00 53 53.6  -72 48 22 &  B2 & 13.96 & 10.79 & 0.07 &-0.03 &   \nl
  90 & 150 & 4034 & 00 53 58.3  -72 28 35 &  B6 & 12.72 & 11.44 & 0.05 & 0.03 &   \nl
  91 & 151 & 1807 & 00 53 59.6  -72 45 53 &  B5 & 12.30 & 10.27 & 0.06 &-0.02 &   \nl
  92 & 152 & 5913 & 00 54  4.2  -72 31 40 &  A3 & 11.87 & 12.23 & 0.03 &-0.06\tablenotemark{c} & \nl
  93 & 153 & 3451 & 00 54  4.0  -72 37 17 &  B8 & 13.58 & 12.51 & 0.06 & 0.01 &   \nl
  94 & 154 & 3095 & 00 54  9.9  -72 41 34 & B0: & 13.55 & 13.30 & 0.05 & 0.26 &   \nl
  95 & 155 & 2975 & 00 54 15.2  -72 42 59 & B0: & 14.34 & 14.54 & 0.10 & 0.30 &   \nl
  96 & 156 & 4384 & 00 54 19.4  -72 17 47 &  A0 & 14.17 & 13.99 & 0.06 & 0.01\tablenotemark{c} & \nl
  97 & 157 & 3938 & 00 54 23.0  -72 17 05 &  B0 & 14.33 & 11.07 & 0.05 & 0.03 &   \nl
  98 & 158 & 3946 & 00 54 23.5  -72 18 59 &  B1 & 14.06 & 10.22 & 0.04 &-0.05 &   \nl
  99 & 160 & 2340 & 00 54 43.5  -72 51 53 &  B0 & 13.99 & 11.34 & 0.07 & 0.07 &   \nl
 100 & 164 & 4070 & 00 55 16.0  -72 41 17 &  B0 & 14.16 & 10.71 & 0.04 & 0.01 &   \nl
 101 & 165 & 3994 & 00 55 19.9  -72 25 59 &  B7 & 12.79 & 11.65 & 0.06 & 0.03 &   \nl
 102 & 169 & 4077 & 00 55 40.3  -72 44 54 &  B1 & 13.90 & 11.07 & 0.05 & 0.03 &   \nl
 103 & 171 & 6141 & 00 55 48.2  -72 39 23 &  B6 & 13.34 & 12.77 & 0.05 & 0.09 &   \nl
 104 & 172 &10309 & 00 55 54.9  -72 08 54 &  B8 & 13.37 & 13.37 & 0.06 & 0.10 &   \nl
 105 & 174 & 8802 & 00 56 38.6  -72 01 36 &  A7 & 12.43 & 14.52 & 0.07 &-0.06\tablenotemark{c} & \nl
 106 & 175 & 4036 & 00 56 38.9  -72 36 30 &  B1 & 13.65 & 11.12 & 0.05 & 0.05 &   \nl
 107 & 176 & 3947 & 00 56 39.7  -72 25 06 &  B1 & 13.93 & 11.10 & 0.04 & 0.03 &   \nl
 108 & 177 & 6454 & 00 56 45.1  -72 03 30 &  O5 & 14.60 & 10.50 & 0.03 & 0.03 &   \nl
 109 & 178 & 3970 & 00 56 48.6  -72 28 42 &  B2 & 14.38 & 11.29 & 0.06 &-0.03 &   \nl
 110 & 180 & 7161 & 00 56 55.8  -72 24 12 &  B5 & 13.19 & 12.27 & 0.09 & 0.07 &   \nl
 111 & 181 &11001 & 00 56 57.8  -72 17 36 &  A0 & 13.83 & 13.15 & 0.04 &-0.03\tablenotemark{c} & \nl
 112 & 182 & 6480 & 00 57  1.6  -72 08 06 &  B0 & 14.33 & 10.81 & 0.03 & 0.00 &   \nl
 113 & 184 & 7143 & 00 57 14.4  -72 22 24 &  B8 & 14.18 & 12.57 & 0.07 &-0.03 &   \nl
 114 & 185 & 6777 & 00 57 24.4  -72 01 30 & B8: & 13.28 & 11.97 & 0.06 & 0.00 &   \nl
 115 & 186 & 4000 & 00 57 27.8  -72 33 06 &  O9 & 14.10 & 10.63 & 0.05 & 0.03 &   \nl
 116 & 189 & 3958 & 00 57 33.3  -72 28 48 &  B1 & 14.51 & 11.21 & 0.05 &-0.01 &   \nl
 117 & 190 &11396 & 00 57 35.1  -72 26 48 &  B6 & 13.52 & 13.19 & 0.08 & 0.11 &   \nl
 118 & 191 & 6482 & 00 57 37.8  -72 13 00 &  B1 & 13.63 & 10.42 & 0.06 & 0.00 &   \nl
 119 & 192 & 6490 & 00 57 38.2  -72 21 49 &  O9 & 14.58 & 11.06 & 0.05 & 0.03 &   \nl
 120 & 193 & 7152 & 00 57 38.1  -72 24 31 &  B1 & 15.40 & 12.54 & 0.09 & 0.03 &   \nl
 121 & 194 & 4008 & 00 57 45.6  -72 35 30 &  B0 & 13.95 & 10.73 & 0.04 & 0.03 &   \nl
 122 & 195 & 5818 & 00 57 45.5  -72 40 12 &  B5 & 13.53 & 14.47 & 0.04 & 0.21 &   \nl
 123 & 196 & 7163 & 00 57 54.7  -72 27 37 &  B1 & 13.93 & 11.57 & 0.09 & 0.07 &   \nl
 124 & 199 & 5274 & 00 58  2.3  -72 35 37 &  A3 & 13.41 & 13.45 & 0.03 &-0.09\tablenotemark{c} & \nl
 125 & 200 & 4025 & 00 58  8.1  -72 38 25 &  A0 & 12.10 & 11.37 & 0.05 &-0.03\tablenotemark{c} & \nl
 126 & 201 & 6479 & 00 58 10.7  -72 10 55 &  B0 & 14.04 & 10.76 & 0.06 & 0.02 &   \nl
 127 & 202 & 6855 & 00 58 15.5  -72 07 25 &  B0 & 14.33 & 11.20 & 0.05 & 0.04 &   \nl
 128 & 204 & 3998 & 00 58 22.5  -72 35 13 & B1: & 14.38 & 11.65 & 0.05 & 0.04 &   \nl
 129 & 205 &11006 & 00 58 23.5  -72 21 31 &  A2 & 12.30 & 13.41 & 0.04 & 0.03\tablenotemark{c} & \nl
 130 & 207 & 6429 & 00 58 33.7  -71 55 43 &  O7 & 14.37 & 10.70 & 0.06 & 0.05 &   \nl
 131 & 208 & 5549 & 00 58 33.7  -72 39 25 &  O9 & 14.10 & 11.49 & 0.10 & 0.10 &   \nl
 132 & 209 & 7135 & 00 58 35.9  -72 24 55 &  B0 & 14.46 & 11.45 & 0.04 & 0.05 &   \nl
 133 & 210 & 6484 & 00 58 36.7  -72 16 19 &  B3 & 12.67 & 10.73 & 0.05 & 0.03 &   \nl
 134 & 211 & 7145 & 00 58 41.8  -72 26 13 &  B8 & 11.50 & 11.67 & 0.07 & 0.11 &   \nl
 135 & 213 & 6607 & 00 58 55.1  -71 56 31 &  A2 & 12.11 & 12.14 & 0.08 &-0.05\tablenotemark{c} & \nl
 136 & 214 & 6964 & 00 58 55.2  -72 13 13 &  B3 & 13.34 & 12.15 & 0.03 & 0.09 &   \nl
 137 & 215 & 3955 & 00 58 55.9  -72 32 01 &  B1 & 12.76 & 10.15 & 0.06 & 0.05 &   \nl
 138 & 217 & 7050 & 00 59  2.3  -72 18 49 & B2: & 14.59 & 11.48 & 0.03 &-0.03 &   \nl
 139 & 218 & 7064 & 00 59  5.2  -72 19 37 &  B1 & 13.80 & 11.29 & 0.05 & 0.06 &   \nl
 140 & 219 & 7010 & 00 59  7.7  -72 16 31 & B2: & 14.50 & 11.23 & 0.06 &-0.04 &   \nl
 141 & 220 & 6446 & 00 59 10.9  -72 05 43 &  O9 & 14.50 & 10.52 & 0.06 &-0.01 &   \nl
 142 & 221 & 3945 & 00 59 10.2  -72 31 31 &  B0 & 13.46 & 11.34 & 0.05 & 0.12 &   \nl
 143 & 222 & 7080 & 00 59 13.5  -72 20 55 &  B3 & 13.20 & 11.42 & 0.06 & 0.04 &   \nl
 144 & 223 & 4007 & 00 59 13.6  -72 38 55 &  B0 & 13.67 & 10.64 & 0.02 & 0.04 &   \nl
 145 & 224 & 6732 & 00 59 16.9  -72 04 37 &  B0 & 14.22 & 11.23 & 0.05 & 0.05 &   \nl
 146 & 225 & 4313 & 00 59 16.8  -72 29 43 &  B8 & 14.05 & 13.07 & 0.03 & 0.02 &   \nl
 147 & 226 & 6481 & 00 59 21.4  -72 17 07 &  B0 & 14.42 & 10.72 & 0.05 &-0.01 &   \nl
 148 & 229 & 8932 & 00 59 27.5  -72 09 49 &  O9 & 11.86 & 10.98 & 0.16 & 0.24 &   \nl
 149 & 230 & 6435 & 00 59 30.1  -72 01 02 &  B2 & 12.75 & 10.64 & 0.05 & 0.05 &   \nl
 150 & 231 &10562 & 00 59 29.6  -72 20 14 &  A0 & 14.20 & 13.54 & 0.04 &-0.03\tablenotemark{c} & \nl
 151 & 232 &   -1 & 00 59 32.9  -72 10 43 &  O9 & 12.36 &  0.00 & 0.00 &-9.99\tablenotemark{e} & \nl
 152 & 234 & 6440 & 00 59 44.3  -72 04 13 &  B2 & 12.98 & 10.75 & 0.04 & 0.04 &   \nl
 153 & 237 & 7019 & 00 59 53.5  -72 19 02 &  B8 & 12.59 & 12.26 & 0.03 & 0.07 &   \nl
 154 & 238 & 6475 & 00 59 56.0  -72 13 32 &  O9 & 13.77 & 10.28 & 0.06 & 0.03 &   \nl
 155 & 240 & 3921 & 01 00  1.1  -72 23 50 &  B3 & 14.03 & 11.42 & 0.05 &-0.02 &   \nl
 156 & 242 & 9303 & 01 00  7.9  -72 13 56 &  B1 & 12.11 & 12.61 & 0.07 & 0.29 &   \nl
 157 & 245 & 6543 & 01 00 16.7  -71 57 02 &  B5 & 13.34 & 11.74 & 0.06 & 0.01 &   \nl
 158 & 250 & 6589 & 01 00 23.8  -71 59 26 &  B9 & 12.82 & 12.71 & 0.08 & 0.06\tablenotemark{c} & \nl
 159 & 251 & 3932 & 01 00 22.9  -72 30 44 &  O7 & 14.75 & 11.09 & 0.06 & 0.05 &   \nl
 160 & 252 & 6450 & 01 00 30.6  -72 10 56 & B2: & 13.07 & 10.63 & 0.05 & 0.03 &   \nl
 161 & 257 & 6486 & 01 00 45.1  -72 23 50 &  B3 & 12.79 & 10.87 & 0.05 & 0.03 &   \nl
 162 & 260 & 6453 & 01 00 52.0  -72 13 32 &  B1 & 13.28 & 10.67 & 0.05 & 0.05 &   \nl
 163 & 263 & 6813 & 01 01  6.9  -72 12 57 &  B6 & 12.85 & 12.07 & 0.04 & 0.07 &   \nl
 164 & 264 & 6422 & 01 01  8.4  -71 59 56 &  B0 & 12.36 &  9.50 & 0.12 & 0.06 &   \nl
 165 & 266 & 6489 & 01 01 10.1  -72 27 26 &  B1 & 12.63 &  9.90 & 0.06 & 0.04 &   \nl
 166 & 267 & 6439 & 01 01 16.0  -72 06 32 &  O8 & 14.92 & 10.72 & 0.03 &-0.01 &   \nl
 167 & 268 & 6448 & 01 01 16.4  -72 12 39 & B0: & 13.14 & 10.42 & 0.06 & 0.07 &   \nl
 168 & 270 & 6477 & 01 01 17.5  -72 17 27 &  B9 & 11.42 & 10.60 & 0.05 & 0.00\tablenotemark{c} & \nl
 169 & 271 & 6476 & 01 01 20.5  -72 17 15 &  B0 & 13.46 & 10.59 & 0.06 & 0.06 &   \nl
 170 & 272 & 9881 & 01 01 23.4  -72 20 09 &  B2 & 14.52 & 12.41 & 0.05 & 0.05 &   \nl
 171 & 273 & 7855 & 01 01 27.9  -72 07 03 &  A1 & 12.18 & 13.32 & 0.04 & 0.07\tablenotemark{c} & \nl
 172 & 274 & 6483 & 01 01 29.8  -72 23 15 &  B0 & 14.02 & 10.60 & 0.06 & 0.01 &   \nl
 173 & 277 & 6824 & 01 01 33.0  -72 14 39 &  B2 & 14.07 & 11.20 & 0.05 &-0.01 &   \nl
 174 & 279 & 6428 & 01 01 35.2  -72 03 03 &  O9 & 14.20 & 10.46 & 0.04 & 0.01 &   \nl
 175 & 280 & 6580 & 01 01 39.9  -72 02 27 &  B0 & 14.66 & 11.74 & 0.05 & 0.05 &   \nl
 176 & 282 & 6447 & 01 01 50.4  -72 13 09 &  O7 & 14.83 & 10.81 & 0.04 & 0.02 &   \nl
 177 & 286 & 6605 & 01 01 57.8  -72 04 15 &  A5 & 12.35 & 13.64 & 0.07 &-0.06\tablenotemark{c} & \nl
 178 & 287 & 6444 & 01 01 57.5  -72 12 39 &  B0 & 12.86 &  9.65 & 0.05 & 0.03 &   \nl
 179 & 290 & 6561 & 01 02  0.8  -72 02 21 &  B1 & 13.93 & 11.49 & 0.07 & 0.06 &   \nl
 180 & 291 & 6816 & 01 02  4.6  -72 15 27 &  B1 & 14.78 & 12.61 & 0.04 & 0.08 &   \nl
 181 & 292 & 6923 & 01 02  4.5  -72 19 03 &  B3 & 13.10 & 11.52 & 0.05 & 0.06 &   \nl
 182 & 296 & 6445 & 01 02  8.8  -72 13 15 &  O8 & 14.38 & 10.77 & 0.03 & 0.04 &   \nl
 183 & 297 & 6419 & 01 02 10.4  -72 00 21 &  B7 & 12.18 & 10.63 & 0.06 & 0.00 &   \nl
 184 & 298 & 6560 & 01 02 12.7  -72 02 51 &  A0 & 12.52 & 12.58 & 0.09 & 0.03\tablenotemark{c} & \nl
 185 & 299 & 6980 & 01 02 13.9  -72 22 09 &  O9 & 14.61 & 12.34 & 0.07 & 0.13 &   \nl
 186 & 300 & 6442 & 01 02 14.8  -72 11 15 &  B0 & 14.46 & 10.69 & 0.05 &-0.01 &   \nl
 187 & 301 & 6512 & 01 02 15.7  -71 59 45 &  B0 & 14.24 & 11.24 & 0.08 & 0.05 &   \nl
 188 & 302 & 9841 & 01 02 19.3  -72 22 03 &  B0 & 14.35 & 14.54 & 0.11 & 0.30 &   \nl
 189 & 303 & 6418 & 01 02 22.3  -72 00 15 &  B1 & 12.81 & 10.16 & 0.07 & 0.05 &   \nl
 190 & 312 & 6441 & 01 02 45.8  -72 12 03 &  B0 & 13.62 &  9.93 & 0.02 &-0.01 &   \nl
 191 & 314 & 8760 & 01 02 48.6  -72 16 39 &  B2 & 12.90 & 10.57 & 0.12 & 0.04 &   \nl
 192 & 315 & 6438 & 01 02 50.0  -72 10 09 &  B8 & 10.92 & 10.82 & 0.07 & 0.09 &   \nl
 193 & 318 & 6436 & 01 02 54.8  -72 09 51 &  O9 & 13.59 & 10.53 & 0.05 & 0.06 &   \nl
 194 & 321 & 6433 & 01 02 57.8  -72 08 03 &  O9 & 13.88 & 10.42 & 0.06 & 0.03 &   \nl

\tablenotetext{a}{No detection by UIT; within 3 arcmin of field edge.}
\tablenotetext{b}{No detection by UIT; \mbfive$>$14.5.}
\tablenotetext{c}{Spectral type later than B8; intrinisic color and E(B$-$V) therefore uncertain.}
\tablenotetext{d}{No magnitude supplied by AV.}
\tablenotetext{e}{Star image is saturated on shortest UIT exposure.
Convolution of UIT B5 bandpass with IUE spectrum SWP22007 gives \mbfive=8.86}

\enddata

\end{deluxetable}
 
\begin{deluxetable}{rrrrrrrrc}

\tablecolumns{9}
\tablecaption{SMC HII Region Ha and FUV Fluxes \label{hiitab}}

\tablehead{
\colhead{\#} & \colhead{DEM} &
\multicolumn{2}{c}{KH \ha} &
\multicolumn{2}{c}{FUV Stellar} &
\multicolumn{2}{c}{FUV Aperture} &
\colhead{``Sky''} \\
\colhead{} & \colhead {\#} &
\colhead{Flux\tablenotemark{a}} &
\colhead{$\sigma$} &
\colhead{Flux\tablenotemark{b}} &
\colhead{$\sigma$} &
\colhead{Flux\tablenotemark{b}} &
\colhead{$\sigma$} &
\colhead{Flux\tablenotemark{c}} \\
 }

\startdata

   1 &  14 &  5.00 &  1.00 & 1.047 & 0.081 & 2.397 & 0.074 & 0.022 \nl
   2 &  15 &  0.70 &  0.30 & 1.067 & 0.101 & 1.665 & 0.061 & 0.015 \nl
   3 &  17 &  1.70 &  0.20 & 0.190 & 0.016 & 0.559 & 0.036 & 0.006 \nl
   4 &  18 &  4.70 &  0.70 & 1.158 & 0.105 & 1.781 & 0.061 & 0.015 \nl
   5 &  20 &  2.90 &  0.50 & 0.755 & 0.048 & 1.987 & 0.063 & 0.015 \nl
   6 &  22 &  2.00 &  0.20 & 0.498 & 0.032 & 1.059 & 0.036 & 0.006 \nl
   7 &  23 &  8.40 &  1.00 & 0.253 & 0.015 & 1.017 & 0.041 & 0.006 \nl
   8 &  24 &  2.80 &  0.30 & 0.146 & 0.010 & 0.747 & 0.036 & 0.006 \nl
   9 &  30 &  2.90 &  0.30 & 0.405 & 0.033 & 1.217 & 0.037 & 0.006 \nl
  10 &  31 &  2.50 &  0.30 & 0.201 & 0.013 & 0.378 & 0.025 & 0.002 \nl
  11 &  32 & 40.00 &  4.00 & 6.202 & 0.463 &11.814 & 0.107 & 0.040 \nl
  12 &  35 &  1.30 &  0.20 & 1.362 & 0.085 & 4.246 & 0.084 & 0.030 \nl
  13 &  36 &  0.90 &  0.20 & 1.793 & 0.108 & 4.655 & 0.085 & 0.030 \nl
  14 &  37 & 10.00 &  1.50 & 1.978 & 0.116 & 4.825 & 0.085 & 0.030 \nl
  15 &  40 &  4.20 &  0.50 & 0.779 & 0.050 & 1.446 & 0.037 & 0.006 \nl
  16 &  42 &  3.20 &  0.40 & 0.364 & 0.024 & 0.943 & 0.036 & 0.006 \nl
  17 &  43 &  6.50 &  0.70 & 0.545 & 0.032 & 1.177 & 0.037 & 0.006 \nl
  18 &  45 & 11.50 &  1.20 & 0.589 & 0.049 & 1.203 & 0.050 & 0.010 \nl
  19 &  46 &  2.30 &  0.30 & 0.720 & 0.033 & 2.779 & 0.034 & 0.022 \nl
  20 &  47 &  1.70 &  0.20 & 0.556 & 0.026 & 1.069 & 0.018 & 0.006 \nl 
  21 &  49 &  5.00 &  1.30 & 1.669 & 0.114 & 3.286 & 0.061 & 0.015 \nl
  22 &  54 & 28.00 &  2.80 & 3.557 & 0.212 & 7.581 & 0.052 & 0.040 \nl
  23 &  55 & 25.00 &  2.50 & 4.585 & 0.217 & 8.356 & 0.047 & 0.040 \nl
  24 &  56 &  4.00 &  0.40 & 3.464 & 0.151 & 5.338 & 0.051 & 0.030 \nl
  25 &  57 &  0.70 &  0.20 & 0.312 & 0.020 & 0.745 & 0.018 & 0.006 \nl
  26 &  63 & 40.00 &  4.00 &22.478 & 1.077 &33.105 & 0.097 & 0.139 \nl
  27 &  69 & 19.50 &  4.00 & 4.353 & 0.217 &13.957 & 0.059 & 0.075 \nl
  28 &  77 &  1.30 &  0.20 & 2.616 & 0.139 & 2.750 & 0.024 & 0.010 \nl
  29 &  80 & 13.50 &  1.50 & 5.744 & 0.326 & 8.266 & 0.044 & 0.089 \nl
  30 &  83 &  0.40 &  0.20 & 0.670 & 0.039 & 0.550 & 0.009 & 0.006 \nl
  31 &  84 &  0.20 &  0.20 & 0.379 & 0.024 & 0.276 & 0.006 & 0.002 \nl
  32 &  85 &  1.30 &  0.30 & 2.562 & 0.156 & 2.369 & 0.022 & 0.030 \nl
  33 &  86 &  1.30 &  0.30 & 2.145 & 0.126 & 2.465 & 0.022 & 0.030 \nl
  34 &  90 &  1.70 &  0.50 & 2.534 & 0.129 & 2.916 & 0.015 & 0.040 \nl
  35 &  93 &  0.80 &  0.20 & 1.575 & 0.091 & 1.552 & 0.016 & 0.015 \nl
  36 &  94 &  1.30 &  0.20 & 1.603 & 0.080 & 1.653 & 0.016 & 0.015 \nl
  37 &  98 &  2.00 &  0.20 & 0.387 & 0.014 & 0.383 & 0.004 & 0.002 \nl
  38 & 100 &  0.70 &  0.20 & 0.299 & 0.017 & 0.353 & 0.005 & 0.006 \nl
  39 & 101 &  0.60 &  0.20 & 0.363 & 0.023 & 0.387 & 0.005 & 0.006 \nl
  40 & 102 &  3.40 &  0.40 & 1.410 & 0.070 & 1.251 & 0.009 & 0.010 \nl 
  41 & 103 & 155.00 & 16.00 &29.339 & 1.747 &93.790 &32.747 & 0.158 \nl
  42 & 111 &  1.20 &  0.20 & 0.521 & 0.037 & 0.491 & 0.006 & 0.006 \nl

\tablenotetext{a}{\dexp{11} \flxl}
\tablenotetext{b}{\dexp{12} \flx}
\tablenotetext{c}{\dexp{12} \flx.  Estimated sky flux contributed by non-SMC
background in the apertures, as described in the text.}

\enddata

\end{deluxetable}

\begin{thebibliography}{DUM}

\bibitem[Azzopardi 1981]{azo} Azzopardi, M. 1981, Ph.D. Thesis, Paul Sabatier
University of Toulose, No. 979. 

\bibitem[AV]{azzvig} Azzopardi, M. \& Vigneau, J. 1982, {\aaps} 50, 291.
                       

\bibitem[Bohlin \ea 1982]{bohlin} Bohlin, R.C., Hill, J.K., Stecher, T.P.,
and Witt, A.N. 1982, {\apj} 255, 87.

\bibitem[Bothun, personal communication]{bothun} Bothun, G. 
(personal communication).


\bibitem[Caplan \ea 1996]{capde} Caplan, J., Ye, T., Deharveng, L., 
Turtle, A., and Kennicutt, R. 1996, {\aap} 307, 403.
 
\bibitem[Charbonnel \ea 1993]{char} Charbonnel, C., Meynet, G., Maeder, A., 
Schaller, G., \& Schaerer, D. 1993, {\aaps} 101, 415.


\bibitem[Cornett \ea 1994]{corn} Cornett, R.H., Hill, J.K., Bohlin, R.C., 
O'Connell, R.W., Roberts, M.S., Smith, A.M., \& Stecher, T.P. 1994, {\apjl}
425, L117.

\bibitem[Cornett \ea 1996]{cornb} Cornett \ea 1996, in preparation.

\bibitem[DEM]{dem} Davies, R.D., Elliott, K.H., and Meaburn, J. 1976, 
{\memras} 81, 89.

\bibitem[Fanelli \ea 1992]{fan} Fanelli, M.N., O'Connell, R.W., Burstein, D. 
\& Wu, C.-C. 1992, {\apjs} 82, 197.

\bibitem[J.K.Hill \ea 1993]{jkha} Hill, J.K., Bohlin, R.C., Cheng, K-P., 
Fanelli, M.N., Hintzen, P.M.N., O'Connell,R.W., Roberts, M.S., Smith, A.M., 
Smith, E.P., \& Stecher, T.P. 1993, {\apj} 413, 610.  

\bibitem[J.K.Hill \ea 1995]{jkhb} Hill, J.K.,  Cheng, K-P., Bohlin, R.C., 
Cornett, R.H., Hintzen, P.M.N., O'Connell,R.W., Roberts, M.S., Smith, A.M., 
Smith, E.P., \& Stecher, T.P. 1995, {\apj} 438, 182.  

\bibitem[R.S.Hill \ea 1994]{rsh} Hill, R.S., Home, A.T., Smith, A.M., 
Bruhweiler, F.C., Cheng, K.-P., Hintzen, P.M., and Oliverson, R.J. 1994, 
{\apj} 430,568.

\bibitem[Hutchings 1982]{hutch} Hutchings, J.B. 1982, {\apj}  255, 70.

\bibitem[KH]{kh} Kennicutt, R.C. Jr. \& Hodge, P.W. 1986, {\apj} 306,130.

\bibitem[Kurucz 1992]{kurucz} Kurucz, R.L. 1992, in {\em The Stellar 
Populations of Galaxies,} ed. B. Barbuy \& A. Renzini (Dordrecht: 
Kluwer Academic), 225. 

\bibitem[Landsman, personal communication]{landsman} Landsman, W.B. 
(personal communication).

\bibitem[Lasker \ea 1989]{lasker} Lasker, B.M., Sturch, C.R., McLean, B.J., 
Russell, J.L., Jenkner, H. \& Shara, M.M. 1989, Space Telescope Science 
Institute Preprint No. 363.

\bibitem[Malumuth \ea 1996]{malumuth} Malumuth, E.M., Waller, W.H., and Parker,
J. Wm. 1996, {\aj} 111, 1128.

\bibitem[Mathewson \ea 1984]{math}Mathewson, D.S., Ford, V.L., Dopita, M.A.,
Tuohy, I.R., Mills, B.Y., \& Turtle, A.J. 1984, {\apjs} 55, 189.

\bibitem[Meyssonier \& Azzopardi 1993]{meyazz} Meyssonier, N. and 
Azzopardi, M. 1993, {\aaps} 102, 451.  

\bibitem[Okumura 1993]{okumura} Okumura, K., 1993, Ph.D. dissertation, 
University of Paris.

\bibitem[Parker \ea 1996]{parker} Parker, J. Wm. 1996, in preparation.

\bibitem[Smith, Cornett, and Hill 1987]{asmith} Smith, A.M., Cornett, 
R.H., and Hill, R.S. 1987, {\apj} 320, 609.

\bibitem[Staveley-Smith \ea 1995]{staveley} Staveley-Smith, L., Sault, R.J., 
McConnell, D., Kesteven, M.J., Hatzidimitriou, D., Freeman, K.C., and 
Dopita, M.A. 1995, {\pasa} 12, 13.

\bibitem[Staveley-Smith \ea 1996]{staveleyb} Staveley-Smith, L., Sault, R.J., 
Hatzidimitriou, D., Kesteven, M.J., \& McConnell, D.
1996, {\mnras} in press. 

\bibitem[Stecher \ea 1992]{stechera} Stecher \ea 1992, {\apjl}  395, L1.
                       
\bibitem[Stecher \ea 1996]{stecherb} Stecher \ea 1996, submitted to 
{\pasp}. 

\bibitem[Stetson 1987]{stetson} Stetson, P.B. 1987, {\pasp} 99, 101.

\bibitem[Waller \ea 1995]{waller} Waller, W.H., Marsh, M., Bohlin, R.C., 
Cornett, R.H., Dixon, W.V., Isensee, J.E., Murthy, J., O'Connell,R.W., 
Roberts, M.S., Smith, A.M., \& Stecher, T.P. 1995 {\aj} 110, 1255.

\bibitem[Westerlund 1990]{westerlund} Westerlund, B.E. 1990 {\aapr} 2,29.
 
\bibitem[Zaritsky 1996]{zar} Zaritsky, D.L. 1996, in preparation.

\end{thebibliography}
\end{document}